\documentclass[twocolumn,prb,aps]{revtex4}
\usepackage[dvips]{graphicx}
\begin{document}
\title{Electron-phonon coupling in the self-consistent 
Born approximation of the $t$-$J$ model}


\begin{abstract} 
We study an undoped $t$-$J$ model with electron-phonon 
interaction using the self-consistent Born approximation 
(SCBA). By neglecting vertex corrections, the SCBA solves 
a boson-holon model, where a holon couples to phonons and 
magnons. Comparison with exact diagonalization results 
for the $t$-$J$ model suggests that the SCBA describes 
the electron-phonon interaction fairly accurately over a 
substantial range of $J/t$ values. Exact diagonalization 
of the boson-holon model shows that the deviations are 
mainly due to the neglect of vertex corrections for small 
$J/t$ and due to the replacement of the $t$-$J$ model by the 
boson-holon model for large $J/t$. For typical values 
of $J/t$, the electron-phonon part $\Sigma_{\rm ep}$ of the 
electron self-energy has comparable contributions from the 
second order diagram in the electron-phonon interaction 
and a phonon induced change of magnon diagrams. A very simple 
approximation to $\Sigma_{\rm ep}$ gives a rather accurate 
effective mass. Using this approximation, we study the factors 
influencing  the electron-phonon interaction. Typically, we 
find that the magnons nominally have a stronger coupling to
the holon than the phonons. The phonons, nevertheless, drive 
the formation of small polarons (self-localization) due to 
important differences between the 
character of the phonon and magnon couplings.

\end{abstract}
\author{O. Gunnarsson and O. R\"osch }
\affiliation{
Max-Planck-Institut f\"ur Festk\"orperforschung, D-70506 Stuttgart, Germany
}

\maketitle

\section{Introduction}\label{sec:1}
There have recently been several experimental indications 
that the electron-phonon interaction plays a substantial 
role for properties of cuprates, for instance in 
photoemission\cite{AL00,KS04,GG04} and neutron scattering 
work.\cite{LP05} The strong effects of the Coulomb interaction 
in the cuprates is often taken into account by using the $t$-$J$ 
model.\cite{FZ88} Including phonons in the $t$-$J$ model,
it was concluded that the Coulomb interaction can enhance 
the effects of the electron-phonon interaction for undoped 
cuprates.\cite{AR92,AM04} It is then interesting to study 
this aspect further. A simple method for treating the undoped 
$t$-$J$ model is the self-consistent Born approximation 
(SCBA),\cite{SS88,CK90,FM91,GM91,ZL92} which can also be 
applied to the $t$-$J$ model with phonons.\cite{AR92,BK96}
This method assumes a quantum N$\rm{\acute e}$el ground-state for the 
undoped system. The excitations of the spin system are 
described by antiferromagnetic magnons. A hole created in, 
e.g., photoemission is assumed to interact with magnons and 
phonons, which are both treated as bosons. This model is 
here referred to as the boson-holon model. The electron 
self-energy of this model is expressed in terms of the 
simplest diagrams, including a boson (magnon or phonon) 
Green's function and a self-consistent electron 
Green's function. Vertex corrections are neglected. 

Here we extend the SCBA study of Ramsak {\it et al.},\cite{AR92} 
focusing on the limit of weak electron-phonon coupling 
for $0.2 \leq J/t\leq 2$.
For strong electron-phonon coupling the SCBA is known to 
break down.\cite{AM04} We first compare results for the 
quasiparticle weight and energy using the SCBA and exact 
diagonalization of the $t$-$J$ model. The results suggest 
that the SCBA describes the electron-phonon interaction 
reasonably well for a substantial range of $J/t$ values,
but that it is less accurate than for the $t$-$J$ model 
without phonons. To trace the sources of errors in the SCBA, 
we use exact diagonalization for the boson-holon model. 
Comparison between these results and results from the SCBA 
shows that for small $J/t(\gtrsim 0.2)$ errors in the SCBA are mainly due 
to the neglect of vertex corrections in the SCBA, while for 
large $J/t$ the main source of errors is the replacement of 
the $t$-$J$ model by the boson-holon model. We then study the 
electron-phonon contributions to the electron self-energy. 
There is a contribution from the diagram containing one 
phonon and one electron Green's function, which for 
noninteracting electrons is the leading contribution. 
Here there is a comparable contribution from diagrams 
containing magnons and one electron Green's  function due 
to the change of the self-consistent 
electron Green's function induced by the electron-phonon
interaction. We also study the effective mass. 
By slightly modifying a previous approach,\cite{AR92} we 
obtain a very simple formula for the effective mass, which 
agrees rather well with exact results within the SCBA. This 
formula is used to illustrate the factors influencing the strength
of the electron-phonon coupling in the undoped $t$-$J$ model. 
We discuss the important difference between the coupling to 
magnons and phonons in terms of strength and effects of vertex 
corrections. We comment on the implications for formation of 
small polarons (self-localization). 

The boson-holon model and the SCBA are described 
in Sec.~\ref{sec:2}. The SCBA results are compared with exact 
diagonalization results for the $t$-$J$ model in Sec.~\ref{sec:3}. 
In Sec.~\ref{sec:4} we compare with exact diagonalization for
the holon-boson model to determine the sources of errors in 
the SCBA. The contributions to the electron-phonon part of the 
electron self-energy and to the
effective mass are discussed in Sec.~\ref{sec:5} and Sec.~\ref{sec:6}, 
respectively. In Sec.~\ref{sec:7} we
compare the coupling to magnons and phonons and discuss 
polaron formation.

\section{Model and method}\label{sec:2}

The $t$-$J$ model \cite{FZ88} is given by the Hamiltonian
\begin{eqnarray}\label{eq:2}
H_{t\textrm{-}J}&=&
J\sum_{<i,j>}\left(
{\bf S}_i\cdot{\bf S}_j-\frac{n_in_j}{4}
\right) \nonumber \\
-&t&\sum_{<i,j>\sigma}(\tilde c_{i\sigma}^{\dagger}
\tilde c_{j\sigma}^{\phantom\dagger}+H.c.),
\end{eqnarray}
where $\tilde c_{i\sigma}^{\dagger}$ creates a  
hole on site $i$ if this site previously had no hole.
The Zhang-Rice singlets are represented by empty sites.
Here $t$ is a  hopping integral, $J$ is the exchange interaction,
${\bf S}_i$ is the spin on site ${\bf R}_i$ and $n_i$ is the 
occupation of site $i$. We use the electron-phonon interaction
\begin{equation}\label{eq:3}
H_{\rm ep}={1\over \sqrt{N}}\sum_{i,{\bf q}} g_{\bf q}(n_i-1)(b_{\bf q}+
b_{-\bf q}^{\dagger})e^{i{\bf q}\cdot {\bf R}_i},
\end{equation}
where $N$ is the number of sites and $b_{\bf q}^{\dagger}$ creates a phonon 
with the wave vector ${\bf q}$. We assume an on-site coupling with 
the strength $g_{\bf q}$. The coupling to hopping integrals and to 
the spin-spin interaction are neglected, since these couplings have 
been found to be weak.\cite{OR04a} In the following we in addition
assume a Holstein type of coupling, i.e., $g_{\bf q}\equiv g$ is 
${\bf q}$-independent.

Following previous work \cite{SS88,CK90,FM91,GM91,ZL92,AR92} for
the undoped system, the Hamiltonian $H_{t\textrm{-}J}+H_{\rm ep}$
is approximately rewritten in terms of a boson-holon model, where 
spinless holons interact with phonons and antiferromagnetic 
magnons, treated as bosons,   
\begin{eqnarray}\label{eq:4}
\tilde H&&= {1\over \sqrt{N}}\sum_{\bf kq}\lbrack h^{\dagger}_{\bf k-q}
h^{\phantom \dagger}_{\bf k}(M_{\bf kq}a_{\bf q}^{\dagger}+
g_{\bf q}b^{\dagger}_{\bf q})+H.c.\rbrack  \nonumber \\
&&+ \sum_{\bf q}(\omega_{\bf q}a_{\bf q}^{\dagger}
a_{\bf q}^{\phantom \dagger}+\omega_{\rm ph}b^{\dagger}_{\bf q}
b^{\phantom \dagger}_{\bf q}),
\end{eqnarray}
where $h^{\dagger}_{\bf k}$ and $a_{\bf q}^{\dagger}$ create 
spinless holons and antiferromagnetic magnons, respectively.
The fermion-magnon coupling is given by
\begin{equation}\label{eq:5}
M_{\bf kq}=\sqrt{8}t\left[\gamma_{\bf k-q}
\sqrt{\nu_{\bf q}^{-1}\!+\!1}
\!-\!\gamma_{\bf k}{\rm sgn}(\gamma_{\bf q})
\sqrt{{\nu_{\bf q}^{-1}}\!-\!1}
\right],
\end{equation}
where $\gamma_{\bf q}=({\rm cos}q_x+{\rm cos}q_y)/2$ and 
$\nu_{\bf q}=(1-\gamma_{\bf q}^2)^{1/2}$. The magnon frequency is
given by $\omega_{\bf q}=2J\nu_{\bf q}$ and the phonon frequency 
by $\omega_{\rm ph}$.

The Hamiltonian $\tilde H$ is treated in the self-consistent Born
approximation. The electron self-energy is then given by 
\cite{SS88,CK90,FM91,GM91,ZL92,AR92}
\begin{eqnarray}\label{eq:6}
\Sigma({\bf k},\omega)&&={1\over N}\sum_{\bf q}\lbrack M_{\bf kq}^2
G({\bf k-q},\omega-\omega_{\bf q})    \\
&& \quad\quad\quad\quad+g_{\bf q}^2G({\bf k-q}
,\omega-\omega_{\rm ph})\rbrack,\nonumber
\end{eqnarray}
where $G({\bf k},\omega)$ is the holon Green's function,
\begin{equation}\label{eq:7}
G({\bf k},\omega)={1\over \omega-\Sigma({\bf k},\omega)}.
\end{equation}
Putting $g_{\bf q}=0$, we obtain the corresponding quantities 
without electron-phonon coupling, $G_0({\bf k},\omega)$
and $\Sigma_0({\bf k},\omega)$. We also introduce the 
electron-phonon part of the electron self-energy
\begin{equation}\label{eq:8}
\Sigma_{\rm ep}({\bf k},\omega)=\Sigma({\bf k},\omega)-\Sigma_0({\bf k},\omega),
\end{equation}
and split $\Sigma_{\rm ep}({\bf k},\omega)$ in two contributions
\begin{equation}\label{eq:9}
\Sigma_{\rm ep}^{\rm 2nd}({\bf k},\omega)={1\over N}\sum_{\bf q}
g_{\bf q}^2G({\bf k-q} ,\omega-\omega_{\rm ph})
\end{equation}
and
\begin{eqnarray}\label{eq:10}
\Sigma_{\rm ep}^{\Delta}({\bf k},\omega)&&=
{1\over N}\sum_{\bf q}\lbrack M_{\bf kq}^2
\lbrack G({\bf k-q},\omega-\omega_{\bf q})\\
&&\quad\quad\quad\quad\quad\ -G_0({\bf k-q} ,\omega-\omega_{\bf q})\rbrack.\nonumber
\end{eqnarray}
Here $\Sigma_{\rm ep}^{\rm 2nd}({\bf k},\omega)$ corresponds to the
second order diagram in the electron-phonon coupling. For 
noninteracting electrons this is the leading contribution 
in $g^2$ to $\Sigma$. For the interacting system there is a 
second contribution of the same order in $g$, $\Sigma_{\rm ep}
^{\Delta} ({\bf k},\omega)$. This is due to the change 
of the contribution from the diagram describing the coupling 
to magnons caused by to the change of the Green's function 
when the electron-phonon coupling is turned on. We also 
introduce\cite {AR92}
\begin{equation}\label{eq:11}
\Sigma_{\rm ep}^{\rm Coh}({\bf k},\omega)={1\over N}\sum_{\bf q}
g_{\bf q}^2G_0^{\rm Coh}({\bf k-q} ,\omega-\omega_{\rm ph}),
\end{equation}
where $G_0^{\rm Coh}({\bf k-q} ,\omega-\omega_{\rm ph})$ only includes 
the coherent part of the Green's function
\begin{equation}\label{eq:12}
G_0^{\rm Coh}({\bf k} ,\omega)={Z_0({\bf k})\over 
\omega-\varepsilon_0({\bf k})}.
\end{equation}
Here $Z_0({\bf k})$ and $\varepsilon_0({\bf k})$ are the 
quasiparticle strength and energy, respectively, in a system 
where $g=0$. Since we consider the limit of weak electron-phonon 
coupling below, we have neglected the effect of the 
electron-phonon coupling on $Z({\bf k})$ and $\varepsilon({\bf k})$ 
in Eqs.~(\ref{eq:11}, \ref{eq:12}). The quasi-particle energy is 
determined by the Dyson equation
\begin{equation}\label{eq:13}
\varepsilon_0({\bf k})=\Sigma_0({\bf k},\varepsilon_0({\bf k})).
\end{equation}
The shift of the quasiparticle energy due to the electron-phonon
interaction is then
\begin{equation}\label{eq:14}
\Delta \varepsilon({\bf k})\equiv \varepsilon({\bf k})-
\varepsilon_0({\bf k})\approx Z_0({\bf k})\Sigma_{\rm ep}({\bf k},
\varepsilon_0({\bf k})),
\end{equation}
where  
\begin{equation}\label{eq:15}
Z_0({\bf k})=\left.\left[1-{\partial \Sigma_0({\bf k},\omega)\over \partial \omega}\right|_
{\omega=\varepsilon_0({\bf k})}\right]^{-1}.
\end{equation}

\section{Comparison with exact diagonalization}\label{sec:3}

\begin{figure}
\centerline{
{\rotatebox{-90}{\resizebox{8.0cm}{!}{\includegraphics {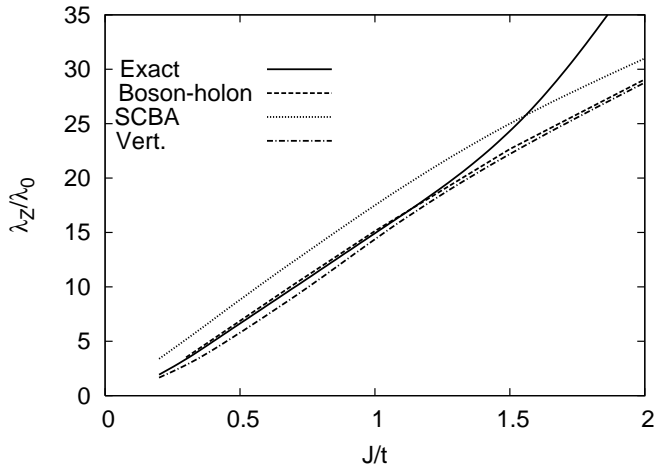}}}}}
\caption{\label{fig:1}$\lambda_Z/\lambda_0$ [Eqs.~(\ref{eq:13a}, 
\ref{eq:13b})] for $\omega_{\rm ph}/t=0.1$ as a function of $J/t$ for a 
$4\times 4$ cluster according to exact diagonalization of the
$t$-$J$ (full line) and the boson-holon (dashed line) models,
the SCBA (dotted line) and SCBA together with the lowest order vertex 
corrections (dash-dotted line) in the limit of a small coupling $g$.
}
\end{figure}

There have been extensive comparisons of results from the SCBA 
and exact diagonalization for small clusters for the case of no 
electron-phonon interaction,\cite{SS88,CK90,FM91,GM91,ZL92} and 
the two methods have been found to agree rather well. Here we 
therefore focus on the changes due to the electron-phonon interaction. 
We define a dimensionless electron-phonon interaction $\lambda_Z$ 
from
\begin{equation}\label{eq:13a}
{Z_0(\pi/2,\pi/2)\over Z(\pi/2,\pi/2)}-1\equiv \lambda_Z,
\end{equation}
where $Z(\pi/2,\pi/2)$ is the quasiparticle weight for 
${\bf k} =(\pi/2,\pi/2)$ including the electron-phonon 
coupling. We study a $4 \times 4$ lattice with periodic boundary conditions in the limit
of weak electron-phonon coupling, for which exact 
diagonalization can easily be performed. Following earlier 
work,\cite{AR92,AM04} we define $\lambda_0$ as the 
corresponding quantity for a single electron at the 
bottom of the band of a two-dimensional Holstein model
with nearest neighbor hopping. Assuming  $\omega_{\rm ph}/t 
\ll 1$ and using a quadratic expansion of the band, we obtain 
\begin{equation}\label{eq:13b}
\lambda_0={g^2 \over 4\pi t \omega_{\rm ph}}.
\end{equation}
We emphasize that by considering the bottom of the band the 
resulting $\lambda_0$ is particularly small.\cite{GS06a}
At larger filling the corresponding $\lambda_0$ is larger
and the resulting enhancement of $\lambda_Z$ is smaller.
It is necessary to pay some extra attention to the ${\bf q}=0$
coupling, $g_0$.  By using Eqs.~({\ref{eq:11}, \ref{eq:12}, 
\ref{eq:14}), we find that in the SCBA this component gives a 
contribution 
\begin{equation}\label{eq:13c}
{1\over N}\left({g_0\over \omega_{\rm ph}}\right)^2Z_0({\bf k})^2
\end{equation}
to $\lambda_Z$.
The ${\bf q}=0$ component just couples to the total number of 
electrons. Since the Green's function describes the addition 
or removal of an electron, we can alternatively calculate the 
exact spectrum for one electron coupling to the ${\bf q}=0$ 
component and then convolute this spectrum with the spectrum 
resulting from the coupling to the ${\bf q}\ne 0$ components. 
The exact ${\bf q}=0$ contribution to $\lambda_Z$ is then
\begin{equation}\label{eq:13d}
{1\over N}\left({g_0\over \omega_{\rm ph}}\right)^2.
\end{equation}
This differs from Eq.~(\ref{eq:13c}) by a factor of $Z_0({\bf k})^2$,
which is typically a very large difference. Although the comparison 
with exact diagonalization can only be done for a small cluster, 
we are primarily interested in infinite systems where the ${\bf q}=0$ 
component plays no role. In discussing $\lambda_Z$ and $\Delta 
\varepsilon$ below we therefore exclude the ${\bf q}=0$ coupling. 

Figure~\ref{fig:1} shows exact results (full line) and results 
from the SCBA (dotted line) for $\omega_{\rm ph}/t=0.1$. The dashed and dash-dotted curve are 
discussed in Sec. \ref{sec:4}. The results agree qualitatively. 
However, quantitatively the agreement is not as good as found 
for the model without electron-phonon interaction. This is, in 
particular, the case for small $J/t$, which are values usually 
assigned to the high-$T_c$ cuprates, and for large values of $J/t$. 
For instance, $\lambda_Z/ \lambda_0$ is about 3.3 and 5.2 according 
to the exact calculation and the SCBA, respectively, for $J/t=0.3$.  
Figure~\ref{fig:2} shows the energy shift $\Delta \varepsilon
(\pi/2,\pi/2)$ [Eq.~(\ref{eq:14})] due to the electron-phonon
interaction for a $4\times 4$ cluster. As in the case of $\lambda_Z$,
the SCBA deviates appreciably from the exact results for small and 
large values of $J/t$. The reason for these deviations are discussed 
in Sec. \ref{sec:4}. The agreement with exact results is still good 
enough to suggest that we can use the SCBA for a qualitative discussion 
of properties of the $t$-$J$ model with phonons.

\begin{figure}
\centerline{
{\rotatebox{-90}{\resizebox{8.0cm}{!}{\includegraphics {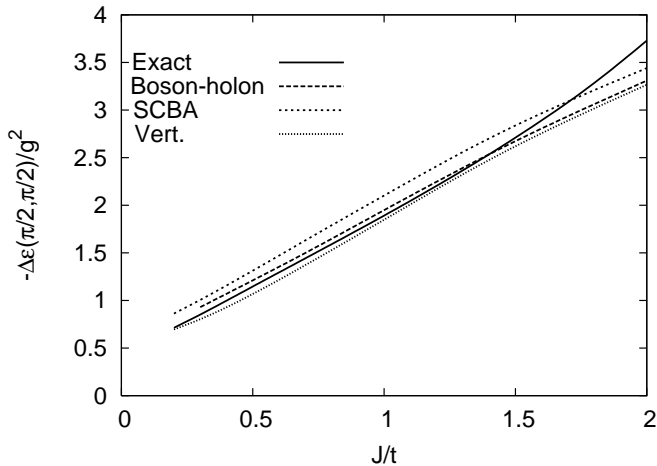}}}}}
\caption{\label{fig:2}Energy shift $\Delta \varepsilon$ for
${\bf k}=(\pi/2,\pi/2)$ [Eq.~(\ref{eq:14})] for a $4\times 4$ 
cluster with $\omega_{\rm ph}/t=0.1$ according to exact diagonalization 
of the $t$-$J$ (full line) and the boson-holon (dashed line) models,
the SCBA (dotted line) and SCBA together with the lowest order vertex 
corrections (dash-dotted line) in the limit of a small coupling $g$.
}
\end{figure}

\section{Accuracy of approximations behind the SCBA}\label{sec:4}

In view of the results in Fig.~\ref{fig:1} and Fig.~\ref{fig:2} it
is interesting to ask for the sources of errors in the SCBA.
We distinguish between two classes of errors: i) the replacement of 
the $t$-$J$ model in Eq.~(\ref{eq:2}) by the boson-holon model
in Eq.~(\ref{eq:4}) and ii) the neglect of vertex corrections 
when solving this model. 

We consider the first class of errors by solving the boson-holon 
model [Eq.~(\ref{eq:4})] using exact diagonalization in the  
limit of weak electron-phonon coupling. The Hilbert space can 
then be limited by only considering states with at most one 
phonon excited. Due to the strong holon-magnon coupling, however, 
it is necessary to consider states with many excited magnons. 
Figures~\ref{fig:1} and \ref{fig:2} compare these results with results 
from diagonalizing the $t$-$J$ model. Except for large $J/t$,
the agreement is very good. This shows that for small and 
intermediate values of $J/t$, the errors in the SCBA are mainly 
due to vertex corrections, while for large $J/t$ the replacement 
of the $t$-$J$ model by the boson-holon model leads to appreciable 
errors.

\begin{figure}
\vskip0.2cm
\rotatebox{0}{\resizebox{4.0cm}{!}{\includegraphics {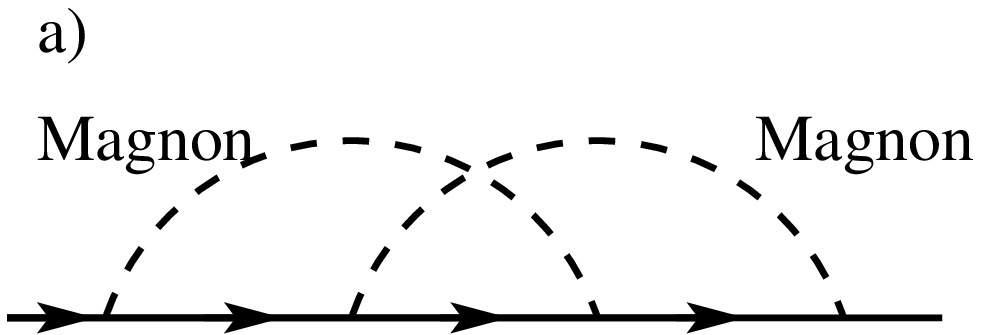}}}
\rotatebox{0}{\resizebox{4.0cm}{!}{\includegraphics {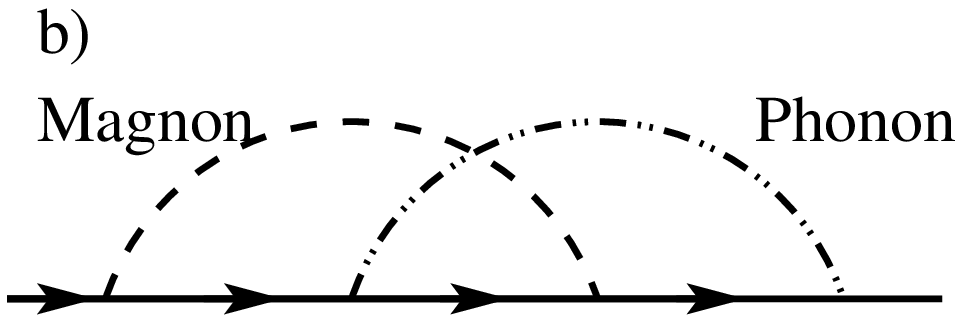}}}
\caption{\label{fig:3}Self-energy diagrams including (a) two crossing 
magnon lines and (b) crossing magnon and phonon lines. The magnon, phonon 
and self-consistent electron propagators are represented by dashed,    
dash-dotted and full lines, respectively.  
}
\end{figure}

We next consider vertex corrections. 
Figure~\ref{fig:3}a shows a second order diagram which is not
included in the SCBA, due to the crossing magnon lines and
resulting vertex correction. It was shown by Liu and Manousakis
\cite{ZL92} that this diagram and many other diagrams neglected
in the SCBA are actually zero due to the symmetry of $M_{\bf kq}$. 
This makes it understandable why the neglect of vertex corrections 
is found to be a rather good approximation for magnons.

In the limit of weak electron-phonon coupling discussed 
in this paper, diagrams with crossing phonon lines do not 
contribute to lowest order in $g_{\bf q}^2$. Diagrams involving 
magnon line(s) crossing one phonon line, however, do contribute 
to this order in $g_{\bf q}^2$. 
Figure~\ref{fig:3}b shows the lowest order diagram of this type. 
In contrast to the pure magnon diagram in Fig.~\ref{fig:3}a,
the diagram in Fig.~\ref{fig:3}b is in general not zero. We have 
included this diagram and an equivalent diagram in the 
calculations, using self-consistent propagators for all 
electron lines in the calculation of $\lambda_Z$ and 
$\Delta \varepsilon$. The results are shown in Figs.~\ref{fig:1} 
and \ref{fig:2} by the dash-dotted lines. For small $J/t$ the 
correction to the SCBA (dashed curve) is large (almost a factor of 
two) and it goes in the correct direction compared with the exact 
result for boson-holon (dashed line) and $t$-$J$ models (full curve). 
Some higher order diagrams are not small, although 
the sum of all higher order diagram apparently almost cancel. The 
rapid convergence for small $J/t$ suggested by Fig.~\ref{fig:1} 
and Fig.~\ref{fig:2} is therefore somewhat misleading. 
We note that these lowest order vertex corrections have a 
substantially smaller effect on the self-energy for larger 
clusters. In these cases, however, it is not possible to 
perform exact diagonalization calculations, and it is therefore not clear
if the SCBA becomes more accurate for large clusters.

We observe that diagrams of the type in Fig.~\ref{fig:3}b 
were neglected in the calculation of the criterion for polaron 
formation.\cite{AM04} If the results in Fig.~\ref{fig:1} 
and Fig.~\ref{fig:2} can be extrapolated to large clusters and 
strong coupling, they suggest that the earlier criterion\cite{AM04} 
for polaron formation may have underestimated the critical $\lambda$. 

We have elsewhere studied vertex corrections to the electron-phonon
interaction for the half-filled Hubbard model in the large $U$ 
limit,\cite{OR05c} which is closely related to the undoped $t$-$J$ model. 
Neglecting vertex corrections and considering  weak electron-phonon 
coupling, we found that a sum rule for the electron-phonon part of 
the imaginary part of the electron self-energy is strongly 
violated,\cite{OR05c} in apparent contradiction to the fairly good 
results found in the SCBA above. The violation of the sum rule 
in the large-$U$ Hubbard model can be traced to the fact that 
the weight of the spectral function only integrates to one half 
over the photoemission energy range. This problem is avoided in 
the SCBA by using the spinless holon Green function, for which 
the spectral function integrates to unity (over the photoemission 
energy range).

\section{Quasiparticle energy}\label{sec:5}

\begin{figure}
\centerline{
{\rotatebox{-90}{\resizebox{7.0cm}{!}{\includegraphics {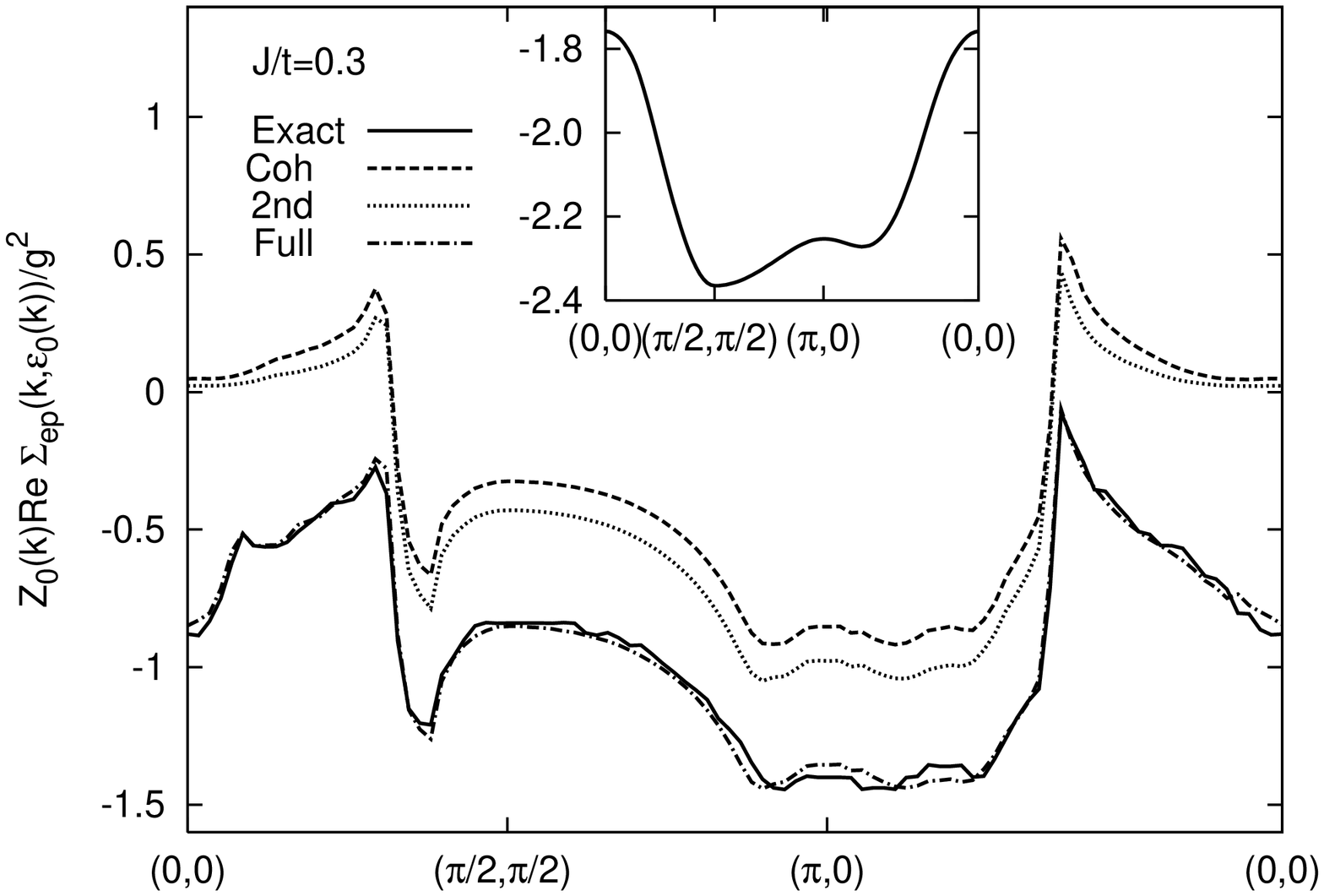}}}}}
\caption{\label{fig:4}$Z_0({\bf k})\Sigma_{\rm ep}({\bf k},
\varepsilon_0({\bf k})$ as a function of ${\bf k}$ along the $(0,0)-
(\pi/2,\pi/2)$, $(\pi/2,\pi/2)-(0,\pi)$ and $(0,\pi)-(0,0)$ directions. 
Result are shown for $\Sigma_{\rm ep}^{\rm Coh}$ [Eq.~(\ref{eq:11}] 
(dashed line), $\Sigma_{\rm ep}^{\rm 2nd}$ [Eq.~(\ref{eq:9})] 
(dotted line) and $\Sigma_{\rm ep}$ [Eq.~(\ref{eq:8})] 
(dash-dotted line) as well as for the exact
$\Delta \varepsilon({\bf k})$  (full line). The inset shows 
the dispersion of $\varepsilon_0({\bf k})$. The parameters are 
$t=1$, $J/t=0.3$, $g/t=0.05$ and $\omega_{\rm ph}/t=0.1$.
}
\end{figure}

In the remainder of the paper we focus on the SCBA and
first analyze the quasiparticle energies. We have performed 
calculations for $96 \times 96$ lattices, using $t=1$, $J/t=0.3$, 
$g/t=0.05$ and $\omega_{\rm ph}/t=0.1$. The self-energy was 
broadened by adding a small imaginary part $\delta/t=0.005-0.01$ 
to the energy. Figure~\ref{fig:4} compares results for $\Delta 
\varepsilon({\bf k})$ (full line), determined from the Dyson equation
using the full self-energy, with $Z_0({\bf k})\Sigma_{\rm ep}
({\bf k},\varepsilon_0({\bf k}))$ [Eq.~(\ref{eq:14})] using three 
approximations for the self-energy. The figure illustrates that 
$\Sigma_{\rm ep}^{\rm Coh}$ (dashed line) is a rather good 
approximation to $\Sigma_{\rm ep}^{\rm 2nd}$ (dotted line), i.e., 
the incoherent part of $G_0$ included in $\Sigma_{\rm ep}^{\rm 2nd}$ 
does not contribute much to the self-energy. It is interesting, 
however, that $\Sigma_{\rm ep}^ {\rm 2nd}$ (dotted line) is a 
rather poor approximation to $\Sigma_{\rm ep}$ (dash-dotted line), 
and it only contributes about half the magnitude for $J/t=0.3$. 
Both $\Sigma_{\rm ep}^ {\rm 2nd}$ and $\Sigma_{\rm ep}^{\Delta}$ 
are of the order $g_{\bf q}^2$. For large values of $J/t$ this 
difference is smaller. For noninteracting electrons $\Sigma_{\rm ep}$ 
is the only contribution of this order, and the interest has therefore 
often focused on this contribution. Figure~\ref{fig:4} hows that 
this is not a good approximation for the present model and $J/t=0.3$. 
The full line and the dash-dotted line differ slightly since the 
solution in Eq.~(\ref{eq:14}) of the Dyson equation is only approximate
for a finite $g$.

To better understand the results for $\Sigma_{\rm ep}^{\rm 2nd}$, we 
notice that for a ${\bf q}$-independent $g_{\bf q}\equiv g$, 
Im $\Sigma_{\rm ep}^{\rm 2nd}$ takes a very simple form 
\begin{equation}\label{eq:16}
{\rm Im} \Sigma_{\rm ep}^{\rm 2nd}({\bf k},\omega)=\pi g^2 
A(\omega-\omega_{\rm ph}),
\end{equation}
where $A(\omega)=\sum_{\bf k} {\rm Im} G_0({\bf k},\omega
-i0^{+})/(N\pi)$ is the ${\bf k}$-averaged spectral function. Figure~\ref{fig:5} shows
$A(\omega)$. Since we used $\omega_{\rm ph}/t=0.1$ in Fig.~\ref{fig:4}, 
the onset of Im $\Sigma_{\rm ep}$ has been shifted by $0.1t$ above the bottom 
of the band. States below this onset are then shifted strongly 
downwards, while states above the onset are shifted less or are 
even shifted upwards. From the inset of Fig.~\ref{fig:4}, we can 
see that states around $(\pi/2,\pi/2)$ and along the line 
$(\pi/2,\pi/2)-(\pi,0)$ are below the onset and are shifted 
strongly downwards, in particular states 
which are just below the onset, while $\Sigma_{\rm ep}^{\rm 2nd}$ 
becomes positive for ${\bf k}$-vectors along the lines $(0,0)-
(\pi/2,\pi/2)$ and $(\pi,0)-(0,0)$ close to $(0,0)$.

In a similar way, we can understand $\Sigma_{\rm ep}^{\Delta}$ 
[Eq.~(\ref{eq:10})], although in this case the coupling 
$M_{\bf k q}$ and the energy $\omega_{\bf q}$ have strong 
${\bf k}$- and ${\bf q}$-dependencies. $\Sigma_{\rm ep}^{\Delta}$
is due to the coupling to the changes of $A({\bf k},\omega)$ 
caused by the electron-phonon coupling. Figure~\ref{fig:5} shows 
the ${\bf k}$-average $A(\omega)$ with (full line) and without 
(dashed line) electron-phonon coupling. The parameters are the 
same as for Fig.~\ref{fig:4}, except that $g/t=0.1$ to enhance 
the effect of the electron-phonon coupling. This difference in 
$A(\omega)$ is positive and particularly large at $\omega 
\approx -2.2$. Since $\omega_{\bf q}$ can be fairly large, 
ranging from zero to $2J=0.6t$, Im $\Sigma_{\rm ep}^{\Delta}$ is 
shifted substantially upwards in frequency. As a result Re 
$\Sigma_{\rm ep}^{\Delta}$ is negative for the whole quasiparticle 
band. 

\section{Effective mass}\label{sec:6}

\begin{figure}
\centerline{
{\rotatebox{-90}{\resizebox{7.5cm}{!}{\includegraphics {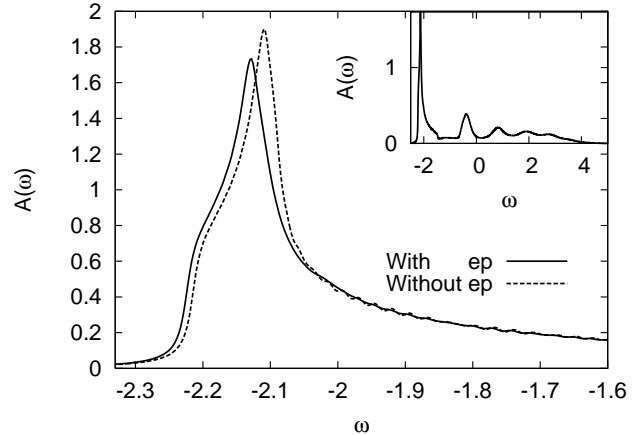}}}}}
\caption{\label{fig:5}Spectral function $A(\omega)$ with (full line)
and without (dashed line) electron-phonon coupling. The parameters
are the same as in Fig.~\ref{fig:4}, except $g=0.1t$. A Lorentzian 
broadening with FWHM $0.2t$ was used. 
}
\end{figure}

We next consider the effective mass, essentially following 
Ramsak {\it et al}.\cite{AR92} Since $\Sigma_{\rm ep}^{\Delta}$ 
is relatively ${\bf k}$-independent, we neglect it and we 
only consider $\Sigma_{\rm ep}^{\rm Coh}$. For simplicity, 
we assume that $Z_0({\bf k})=Z_0(\pi/2,\pi/2)$ is 
${\bf k}$-independent. We furthermore approximate the quasiparticle
dispersion by assuming that it can be expanded quadratically
around the four minima $(\pm\pi/2,\pm\pi/2)$. Two masses are
introduced, $m_{\parallel}$ and $m_{\perp}$, which describe 
the dispersion parallel and perpendicular to the $(0,0)-(\pi,\pi)$ 
direction, respectively. The summation 
over the Brillouin zone in Eq.~(\ref{eq:11}) is replaced by an 
integration over all of ${\bf q}$-space, assuming that contributions 
far away from $(\pm\pi/2,\pm\pi/2)$ are small because of the large 
energy denominator in Eqs.~(\ref{eq:11}, \ref{eq:12}). By using the 
solution of the Dyson equation Eq.~(\ref{eq:14}), we then obtain
\begin{equation}\label{eq:17}
\Delta \varepsilon({\bf k})=4\left({1\over 2\pi}\right)^2\int d^2q {g^2Z_0^2\over 
\varepsilon_0({\bf k})-\varepsilon_0({\bf k-q})-\omega_{\rm ph}},
\end{equation}
where the factor four is due to the presence of four equivalent minima
$(\pm \pi/2,\pm \pi/2)$. One factor of $Z_0$ comes from the Green's 
function and a second factor from solving the Dyson equation. Defining 
the effective mass along the parallel direction as $1/m^{\ast}=d^2 
\varepsilon({\bf k})/dk_{\parallel}^2$, we obtain
\begin{equation}\label{eq:18}
{m_{\parallel}\over m^{\ast}}-1=-
{2g^2Z_0^2 \sqrt{m_{\parallel}m_{\perp}} \over \pi\omega_{\rm ph}}
\equiv \left({1\over 1+\lambda_m}-1\right)
\end{equation}
where the second equality defines the electron-phonon coupling 
$\lambda_m$. Focusing on large $J/t$, Ramsak {\it et al.}\cite{AR92} 
obtained the same result except for a factor $Z_0$ resulting from the 
Dyson equation (\ref{eq:14}). Without this factor the rather
good agreement with the exact result in Fig.~\ref{fig:6}
would be lost for small $J/t$. $\lambda_m$ is compared with the
corresponding quantity for the Holstein model,\cite{AR92} 
which is identical to the $\lambda_0$ defined via $Z$ in 
Eq.~(\ref{eq:13b}). 

Figure~\ref{fig:6} shows results for 
$\lambda_m/\lambda_0$ as a function of $J/t$, using 
the second derivative of the exact $\varepsilon({\bf k})$ 
(full line) and of $\varepsilon({\bf k})$ obtained from 
$\Sigma_{el}^{\rm Coh}$ (dotted line) as well as Eq.~(\ref{eq:18}) 
(dashed line). $\lambda_m/\lambda_0$ is different from 
$\lambda_Z/\lambda_0$ in Fig.~\ref{fig:1}. The main reason for 
this difference is that Fig.~\ref{fig:1} shows result for a 
$4 \times 4$ cluster while Fig.~\ref{fig:6} shows result for 
a large cluster ($96 \times 96$ or $128\times 128$), but the 
two quantities are somewhat different also for identical 
clusters. The results based on $\Sigma_{el}^{\rm Coh}$ (dotted 
line) agree rather well with the exacts result for intermediate 
and large values of $J/t$, while they are too small for small 
values of $J/t$. The deviation for small $J/t$ is primarily due 
to the neglect of $\Sigma_{\rm ep}^{\Delta}$. This term is 
particularly important for small $J/t$, since the magnon energy 
entering in Eq.~(\ref{eq:10}) is proportional to $J$. For large 
$J/t$, the deviation is mainly due to the neglect of the incoherent 
part of $G$. Eq.~(\ref{eq:18}) (dashed curve), which is an 
approximation to the dotted curve, gives a larger $\lambda_m$ 
and it agrees better with the exact result. For small $J/t$, 
the increase is primarily due to neglect of ${\bf k}$-dependence 
of $Z_0({\bf k})$ in the Dyson equation when deriving Eq.~(\ref{eq:18}). 
For large large $J/t$ the increase is due to several small errors 
in the approximations. The coupling $g/t=0.1$ is somewhat too
large to give the weak-coupling limit, in particular for large 
$J/t$. 

\begin{figure}
\centerline{
{\rotatebox{-90}{\resizebox{7.5cm}{!}{\includegraphics {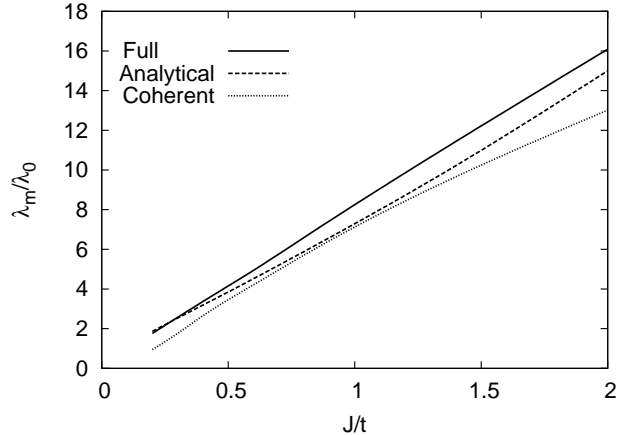}}}}}
\caption{\label{fig:6}$\lambda_m/\lambda_0$ as a function of $J/t$
calculated from the quasiparticle energy $\varepsilon({\bf k})$ 
(full line), Eq.~(\ref{eq:18}) (dashed line) and using the self-energy 
$\Sigma^{\rm Coh}_{\rm ep}$ [Eq.~(\ref{eq:11})]. The parameters are 
the same as in Fig.~\ref{fig:4}, except that $g=0.1t$ and $J/t$ is varied. 
}
\end{figure}

We are now in the position to interpret the enhancement 
of $\lambda_m$ compared with $\lambda_0$ obtained at the 
bottom of the band for a Holstein model. Equation~(\ref{eq:18}) 
contains a factor $Z_0^2$ which tends to reduce the coupling
due to the transfer of spectral weight far away from the energies 
studied.  On the other hand, the factor $\sqrt{m_{\parallel} 
m_{\perp}}$ describes how the energy denominator is reduced 
by the large effective masses, bringing spectral weight 
closer to the relevant energies.\cite{AR92} This is an effect 
of correlation and antiferromagnetism, and it is important 
for the enhancement of the electron-phonon interaction in 
this model. In addition, there is a factor of four resulting 
from the presence of the four equivalent minima $(\pm \pi/2,
\pm \pi/2)$. For $J/t=0.2$ we find that $Z_0^2=0.05$ and 
$\sqrt{m_{\parallel}m_{\perp}}=10m_0$, where $m_0=1/(2t)$ 
is the mass at the bottom of the band in the Holstein model. 
In this case the factor four from the equivalent minima is 
crucial, since the electron-phonon interaction would otherwise 
have been suppressed in the $t$-$J$ model, while now it is 
enhanced by a factor of 1.8. For $J/t=2$, we obtain $Z_0^2=0.57$ and 
$\sqrt{m_{\parallel} m_{\perp}}=6.5m_0$, giving the enhancement 16
(15 according to Eq.~(\ref{eq:18})). In this case, the 
large mass plays a crucial role for the enhancement of the 
electron-phonon interaction. We notice, however, that the comparison
here has been done with $\lambda_0$ calculated for a single 
electron at the bottom of the band of a Holstein model. Had the
comparison been made with a half-filled Holstein model, the result 
would have been a smaller enhancement or no enhancement at all.\cite{GS06a}
By starting from the three-band model and studying the half-breathing
phonon, however, it is found that the coupling constants $g_{\bf q}$
are enhanced by correlation effects.\cite{OR04c}

\section{Comparison of coupling to magnons and phonons}\label{sec:7}

We define an average dimensionless coupling constant for the magnons
\begin{equation}\label{eq:19}
\lambda_M\equiv {1\over N}\sum_{\bf k}\lambda_{M{\bf k}}
={1\over N^2}\sum_{\bf q k}{2M_{\bf kq}^2
\over 8t \omega_{\bf q}}={t\over 2J},
\end{equation}
where $\lambda_{M (\pi/2,\pi/2)}=0.65t/J$. For La$_2$CuO$_4$ 
the corresponding quantity due to phonons is $\lambda=
1.2$.\cite{OR05b} For a typical value $J/t=0.3$,\cite{JJ00} 
the coupling to magnons, $\lambda_M=1.67$ and $\lambda_{M 
(\pi/2,\pi/2)}=2.2$, is  stronger than the coupling to phonons. 
It might then seem that the coupling to magnons is more important 
for the experimentally observed \cite{KS04} polaron formation in 
undoped cuprates. This is, however, misleading. The value 
of $\lambda$ needed for formation of small polarons is reduced with the boson 
frequency.\cite{HF97,SC97,AA99} This somewhat favors phonons, since 
they typically have lower frequencies than the magnon frequencies of 
the order of $J$. To see the main difference, however, it is 
necessary to consider vertex corrections.  

To describe the formation of small polarons due to phonons, it 
is crucial to go beyond the SCBA, since vertex corrections
including phonon propagators become very important in the 
strong-coupling limit.\cite{AM04} Actually, if these vertex 
corrections are neglected, polaron formation is not properly 
obtained.\cite{AM04} On the other hand, it has been argued that 
vertex corrections including magnon propagators are not very 
important in the $t$-$J$ model.\cite{SS88,CK90,GM91,ZL92} 
As discussed in Sec.~\ref{sec:4}, the lowest order vertex 
correction in Fig.~\ref{fig:3}a is identically zero due to 
the symmetry of $M_{\bf kq}$, and classes of higher order 
vertex corrections are also zero.\cite{ZL92} If we assume a 
Holstein type of electron-phonon coupling, however, there are 
no similar arguments for diagrams with crossing phonon lines 
being zero. This explains why the holon-phonon interaction, 
but not the holon-magnon interaction, leads to polaron formation. 

\section{Summary}\label{sec:8}
 
To summarize, we have studied the self-consistent Born approximation
(SCBA) in the limit of weak electron-phonon coupling. While the SCBA 
has been shown to be quite accurate for a pure $t$-$J$ model, we find 
that it is less accurate when the electron-phonon interaction is 
included. To study the reason for this, we performed exact diagonalization
calculations for the boson-holon model. Comparing the results with 
the SCBA results, we find that the main errors of the SCBA are due 
to the neglect of vertex corrections for small $J/t$ and due to the     
the introduction of the boson-holon model itself for large $J/t$. 
Studying the electron-phonon part of the self-energy, we find that 
in addition to the second order term $\Sigma_{\rm ep}^{\rm 2nd}$ 
known from the theory of noninteracting electrons, there is a second 
term of the same order, $\Sigma_{\rm ep}^{\Delta}$. For $J/t \approx 0.3$ 
this term makes a similar contribution to $\Sigma_{\rm ep}$ as 
$\Sigma_{\rm ep}^{\rm 2nd}$. We have shown that a very simple derivation 
of the effective mass gives a rather accurate result, illustrating    
the factors enhancing and suppressing the electron-phonon coupling. 
The coupling to magnons can be considered stronger than the coupling 
to phonons for realistic parameters.  Nevertheless, the phonons drive 
the formation of small polarons for undoped cuprates, due to the difference 
between phonons and magnons, in particular the different importance 
of vertex corrections.

\end{document}